\documentclass[twocolumn]{aastex631}
\usepackage{multirow} 
\usepackage{amsmath}

\definecolor{celticsgreen}{rgb}{0, 0.478, 0.2}

\reportnum{MIT-CTP/5784}

\begin{document}

\title{Domain adaptation in application to gravitational lens finding}

\correspondingauthor{Hanna Parul}
\email{hparul@crimson.ua.edu}

\author[0000-0000-0000-0000]{Hanna Parul}
\affiliation{Department of Physics \& Astronomy, University of Alabama, Tuscaloosa, AL 35401, USA}

\author[0000-0002-6222-8102]{Sergei Gleyzer}
\affiliation{Department of Physics \& Astronomy, University of Alabama, Tuscaloosa, AL 35401, USA}


\author[0000-0002-6609-3495]{Pranath Reddy}
\affiliation{University of Florida, Gainesville, FL 32611, USA}

\author[0000-0003-1205-4033]{Michael W.~Toomey}
\address{Center for Theoretical Physics, Massachusetts Institute of Technology, Cambridge, MA 02139, USA}



\begin{abstract}
The next decade is expected to see a tenfold increase in the number of strong gravitational lenses, driven by new wide-field imaging surveys. To discover these rare objects, efficient automated detection methods need to be developed. In this work, we assess the performance of three domain adaptation techniques -- Adversarial Discriminative Domain Adaptation (ADDA), Wasserstein Distance Guided Representation Learning (WDGRL), and Supervised Domain Adaptation (SDA) -- in enhancing lens-finding algorithms trained on simulated data when applied to observations from the Hyper Suprime-Cam Subaru Strategic Program. We find that WDGRL combined with an ENN-based encoder provides the best performance in an unsupervised setting and that supervised domain adaptation is able to enhance the model's ability to distinguish between lenses and common similar-looking false positives, such as spiral galaxies, which is crucial for future lens surveys. 
\end{abstract}



\section{Introduction} \label{sec:intro}

Gravitational lensing, the distortion of light from a distant source due to the presence of a massive object between that source and the observer, is one of the most important phenomena predicted by the general relativity theory and a powerful tool to study the universe. 
For example, the sensitivity of lensing observables to the distribution of mass in the lens provides a way to study the dark matter profiles of individual galaxies, galaxy groups, and galaxy clusters \citep{Koopmans2009, Barnabe2011, Sonnenfeld2013, Newman2015, Li2018, Etherington23}. It also allows to probe dark matter substructures on subgalactic scales \citep{Daylan2018, Brehmer2019, DiazRivero20, Bayer2023}, and to test various dark matter theories \citep{Li2016, Gilman2018, Alexander20, Sengul22, Vegetti23, AnauMontel23, Gilman23}.
Time-delay measurements of gravitationally lensed time-variable sources, such as supernovae and quasars, allow for measuring absolute distances and constraining the Hubble constant and other cosmological parameters \citep{Refsdal, Shajib20, Wong20, Treu2022, Birrer22, Suyu24}. 
Cluster-scale lenses provide magnification that enables spectroscopic 
studies of high-redshift galaxies, helping to uncover star formation processes in the early universe \citep{Rigby2018, Sharon2022, Zhang2023, Mainali2023, Keerthi2024}. 

Strong gravitational lensing, characterized by the occurence of multiple images, arcs, and rings, requires precise alignment between foreground and background sources. Therefore, these systems are very rare, and their discovery demands extensive search efforts. 
Upcoming wide-field surveys, such as Euclid Wide Survey \citep{Euclid, EuclidWide} and Legacy Survey of Space and Time \citep[LSST;][]{LSST}
are expected to deliver on the order of $10^5$ strong lenses \citep{Collett2015}. However, finding them will require searching through hundreds of millions of sources, which is infeasible for human inspection. Over the last 20 years, various methods have been employed to discover gravitational lenses, such as citizen science \citep[e.g.][]{Marshall2016}, spectroscopic selection of lens candidates \citep{Bolton2004, Bolton2006, Shu2017}, arcs and ring-like feature finders \citep{More2012, Gavazzi2014, Sonnenfeld2018}.
Nevertheless, in recent years, the majority of lens discoveries have come from deep learning algorithms.

Typically, the lens finding problem is treated as an image classification problem, which is well-suited for Convolutional Neural Networks (CNN) \citep{Lecun1998}. 
Supervised CNN-based algorithms have demonstrated superior performance in the gravitational lens finding challenge \citep{metcalf_challenge} and have recently been applied to various wide-field surveys, yielding a few thousand lens candidates \citep[e.g.][]{petrillo, Pourrahmani, Canameras_HSC, Canameras_panstarrs, Li, Rojas, Shu, Storfer, Huang2020, Huang2021, Jacobs2019}.
To ensure that the model, trained in a supervised way, is able to learn relevant features and generalize well to unseen data, the training dataset typically requires a large number of labeled samples (on the order of $10^3 - 10^4$), which exceeds the number of known lenses. For that reason, it is common to use simulated lenses at the training stage (however, see e.g. \citealp{Huang2020, Huang2021} for supervised training only with observational data).

Despite efforts to increase the realism of simulations, differences between real observations and mock images are unavoidable and can significantly degrade the performance of the model \citep{Ciprijanovic20, Ciprijanovic22}.
Among the ways to tackle this issue are, for example, implementing unsupervised methods and train a model purely on observational data, as in \citet{Cheng20} and \citet{Stein22}, or applying domain adaptation techniques to reduce the gap between the simulated and observational domains.

Domain adaptation (DA) is a class of methods applied in situations when the training and test datasets come from different, but related domains, namely the \textit{source} and the \textit{target} domains. Based on the availability of labels in the target dataset, DA methods can be divided into three categories: unsupervised domain adaptation (UDA), when ground truth for the target data is unknown; semi-supervised domain adaptation (SSDA), where labels are available for a fraction of the target data; and supervised domain adaptation (SDA), where labels for the entire dataset are accessible. In astrophysics, domain adaptation has found various applications, including, but not limited to: improving the performance of galactic morphology classifiers across different surveys \citep{Xu2023_DA, Ciprijanovic23}, identifying merging galaxies in simulated and observational datasets \citep{Ciprijanovic21}, and classifying strong gravitational lenses with different dark matter substructure simulated for surveys with varying observational characteristics \citep{deeplense_DA}.

In this paper, we assess how much domain adaptation techniques can enhance the performance of lens-finding algorithms trained in a supervised way on simulated or real data. Building upon the work in \citet{deeplense_DA}, we examine three domain adaptation methods: Adversarial Discriminative Domain Adaptation (ADDA), Wasserstein Distance Guided Representation Learning (WDGRL), and Supervised Domain Adaptation (SDA). The paper is structured as follows: Section \ref{DA} describes the algorithms we employ; Section \ref{sec:data} presents the datasets used in the experiments; and Section \ref{arch} outlines the neural network architecture and training process. We discuss our results in Section \ref{res} and conclude in Section \ref{disc}.

\section{Domain Adaptation methods}\label{DA}
In the context of gravitational lens finding, our goal is to train a model on a source dataset containing simulated lenses and adapt it to a smaller unlabeled target dataset with real lenses, using either the Adversarial Discriminative Domain Adaptation (ADDA) or the Wasserstein Distance Guided Representation Learning (WDGRL) method. 
We will also explore the supervised setting for domain adaptation and compare the supervised classifier trained purely on real data with the Supervised Domain Adaptation (SDA) algorithm trained on a labeled target dataset. 
A brief description of each domain adaptation algorithm follows.
For the remainder of the section, $(\mathbf{X}^s, Y^s)$ denotes images and labels of the source dataset, while $(\mathbf{X}^t, Y^t)$ represents images and labels of the target dataset.

\subsection{Adversarial Discriminative Domain Adaptation}
The ADDA framework \citep{adda} includes source and target encoders $M_s$ and $M_t$, respectively, which map the input images to a lower-dimensional latent space, discriminator $D$, that learns to identify whether the input representation comes from the source or target domain, and classifiers $C_s$ and $C_t$, which act on representations and output class labels for source and target data, respectively. The goal of the training process is to minimize the distance between the source and target representations, $M_s(\mathbf{X}^s)$ and $M_t(\mathbf{X}^t)$. In this case, the source classifier $C_s$, trained on source data in a supervised manner, could be directly applied to the target data, such that $C_t = C_s$.

The minimization of the distance between representations is achieved through the alternating optimization of the discriminator and the target encoder. The discriminator is updated to minimize the discriminator loss:
\begin{equation}
    \begin{split}
         & \mathcal{L}_{\mathrm{adv}_{D}} (\mathbf{X}^s, \mathbf{X}^t, M_s, M_t) = \\ 
         & - \mathbb{E}_{\mathbf{x}^s \sim \mathbf{X}^s} [\log D(M_s(\mathbf{x}^s))] - \mathbb{E}_{\mathbf{x}^t \sim \mathbf{X}^t} [\log (1 - D(M_t(\mathbf{x}^t)))]
    \end{split}
\end{equation}
The target encoder learns to fool the discriminator by finding representations of the target images that are indistinguishable from the representations of the source images. It is optimized according to a standard loss function with inverted labels:
\begin{equation}
    \mathcal{L}(\mathbf{X}^t, D) = \mathbb{E}_{\mathbf{x}^t\sim\mathbf{X}^t} [\log D(M_t(\mathbf{x}^t))]
\end{equation}

\subsection{Wasserstein Distance Guided Representation Learning}

The WDGRL method \citep{wdgrl} is inspired by the idea of Wasserstein Generative Adversarial Networks (WGAN). WGANs \citep{wgan} were introduced to address the problem of standard GANs when the discriminator quickly learns to distinguish between fake and real samples and ceases to provide reliable gradient information. The employment of the Wasserstein distance allowed for obtaining more stable gradients even when the two distributions are distant. 

The Wasserstein distance is a measure of similarity between two probability distributions.
For the case of the first Wasserstein distance, it can be written in the following form:
\begin{equation}
    W_1(\mathbb{P}, \mathbb{Q}) = \sup_{||f_w||_L \leq 1} \mathbb{E}_{x \sim \mathbb{P}}[f(x)] - \mathbb{E}_{x\sim \mathbb{Q}}[f(x)]
\end{equation}
where $\mathbb{P}$ and $\mathbb{Q}$ are two probability measures and $f$ is 1-Lipschitz function. 

WDGRL applies this concept to learn domain-invariant features through the iterative training of the following components: a domain critic $D$, a feature extractor $M$, and a classifier $C$ with parameters $\theta_D$, $\theta_M$, $\theta_C$, respectively.

The goal of the domain critic is to approximate the Wasserstein distance between the representations by maximizing the domain critic loss $\mathcal{L}_{wd}$:
\begin{equation}
    \mathcal{L}_{wd}(\mathbf{x}^s, \mathbf{x}^t) = \frac{1}{n^s} \sum_{\mathbf{x}^s \in \mathbf{X}^s} D(M(\mathbf{x}^s)) - \frac{1}{n^t} \sum_{\mathbf{x}^t \in \mathbf{X}^t} D(M(\mathbf{x}^t))
\end{equation}
where $n^s$ and $n^t$ are the number of source and target samples, respectively.

To ensure that $D$ is a 1-Lipschitz continuous function, a gradient penalty is enforced on the domain critic parameters:
\begin{equation}
    \mathcal{L}_{grad}(\hat{h}) = (||\nabla_{\hat{h}} D(\hat{h})||_2 - 1)^2,
\end{equation}
where $\hat{h}$ are the points sampled between the source and target representations.

During training, for each mini-batch, the domain critic is first trained to optimality by maximizing the domain critic loss $\mathcal{L}_{wd} - \gamma\mathcal{L}_{grad}$, where $\gamma$ is a balancing coefficient. In practice, the number of iterations to train the domain critic is set by the parameter \texttt{n\_critic}. Next, the optimal domain critic parameters are fixed, and the feature extractor network is trained to minimize the Wasserstein distance and learn invariant representations. Since the learned representations should be informative enough for the classifier $C$, the source labels are incorporated, and the classifier is optimized by minimizing the cross-entropy loss $\mathcal{L}_c(\mathbf{x}^s, y^s)$, where $y^s$ represents the source labels. Therefore, the combined objective function for this step is $\displaystyle \min_{\theta_M, \theta_C} \bigl\{\mathcal{L}_c + \lambda \max_{\theta_D} \mathcal{L}_{wd} \bigr\}$, where $\lambda$ is a trade-off parameter.

\subsection{Supervised Domain Adaptation}
Typically, unsupervised DA methods expect a large amount of target data to be effective. For a limited size target dataset with available labels, supervised DA methods tend to be more powerful.

The SDA approach \citep{sda} aims to align the source and target representations in the embedding space, similar to the unsupervised methods described above. However, access to target labels enables contrastive semantic alignment. This means that the encoder $M$ maps samples from the different domains but with the same class label near to each other in the embedding space, while keeping samples with different class label well separated. 

The alignment of the same-class samples is formulated via minimizing the semantic alignment loss:
\begin{equation}
    \mathcal{L}_{SA} = \sum_{a=1}^{N}d(M(\mathbf{X}_a^s), M(\mathbf{X}_a^t)),
\end{equation}
where $N$ is the number of class labels, $\mathbf{X}_a^s = \mathbf{X}^s|{Y^s=a}$ and $\mathbf{X}_a^t = \mathbf{X}^t|{Y^t=a}$ are conditional random variables, and $d$ is the distance metric in the embedding space. 

To maximize the distance between the representations of samples from different domains and different labels, the separation loss is included:
\begin{equation}
    \mathcal{L}_S = \sum_{a, b|a\neq b} k(M(\mathbf{X}_a^s), M(\mathbf{X}_a^t)),
\end{equation}
where $k$ is the similarity metric between the distributions of representations, which penalizes them if they come close.
Following the implementation in \citet{sda}, we compute the distance, $d$, and similarity, $k$, as average pairwise distances and similarities between points in the embedding space, where for each pair of points they are defined as:
\begin{equation}
        d(M(x_i^s), M(x_j^t)) = \frac{1}{2}\|M(x_i^s) - M(x_j^t)\|^2 
\end{equation}
\begin{equation}
        k(M(x_i^s), M(x_j^t)) = \frac{1}{2}\max(0, m - \|M(x_i^s) - M(x_j^t)\|)^2,
\end{equation}
where margin $m$ is a hyperparameter that defines the minimum desired separation between samples from different classes in the embedding space.

The model also includes a classifier $C$ trained purely on the source dataset in a supervised manner. Therefore, the combined classification and contrastive semantic alignment loss takes the following form:
\begin{equation}
    \mathcal{L}_{CCSA} = (1-\alpha)\mathcal{L}_C + \alpha(\mathcal{L}_{SA} + \mathcal{L}_S),
\end{equation}
where $\mathcal{L}_C$ is a classifier loss and $\alpha$ is a balancing coefficient.

\section{Data} \label{sec:data}
As a source dataset, we use a mixture of simulated lenses and real non-lensed galaxies. For the target dataset, we use purely observational data, combining previously discovered lens candidates with non-lensed galaxies. We describe the construction of these datasets below.

We begin by selecting a parent dataset from Hyper Suprime-Cam Subaru Strategic Program (HSC-SSP) PDR2 Wide field \citep{hscAihara}, following the criteria in \citet{Shu2022} and \citet{Canameras2021}. The selection targets the extended sources brighter than 26 mag in the \textit{g}, \textit{r}, and \textit{i} bands
with the following color cuts: $0.6< \texttt{g\_cmodel\_mag} - \texttt{r\_cmodel\_mag} < 3$, $2< \texttt{g\_cmodel\_mag} - \texttt{i\_cmodel\_mag} < 5$. The color cuts serve to narrow the sample to preferentially red galaxies, as massive red ellipticals are expected to dominate the lens population. We cross-matched the parent dataset with the list of known lens candidates and removed matches with separation less than 40 arcsec to reduce  possible contamination among the non-lensed sources. We then used this cleaned parent dataset to draw negative examples (non-lenses) for both the source and target dataset.
While there is a possibility of undiscovered lenses in the parent sample, we expect the fraction of such serendipitous lenses to be negligible. 

\subsection{Mock lenses}
The essential components for creating a mock gravitational lens are the light profiles of the foreground and background galaxies and the gravitational potential of the lens. To enhance the realism of the mock systems, we employ images of real galaxies for both the foreground and background sources, following the approach in \citet{Canameras_panstarrs}. 

For the deflectors (or foreground galaxies) we selected galaxies from the parent dataset that also have spectroscopic redshift and velocity dispersion measurements in the Sloan Digital Sky Survey \citep[SDSS;][]{SDSS}. We used the HSC Cutout service\footnote{\url{https://hsc-release.mtk.nao.ac.jp/das_cutout/pdr2/}} to extract 72 x 72 pixel cutouts which corresponds to 11.5 x 11.5 arcseconds in the \textit{g}, \textit{r}, and \textit{i} filters.

For the background sources, we used a sample of galaxies observed in the Hubble eXtra Deep Field \citep{xdf} in the F435W, F606W, and F775W bands with redshifts from \cite{Inami}. 
All the {\it HST} data used in this paper can be found in MAST: \dataset[10.17909/T9RG6J]{http://dx.doi.org/10.17909/T9RG6J}. 

To simulate strong galaxy-galaxy lens systems, we used the package \texttt{lenstronomy}\footnote{\url{https://github.com/lenstronomy/lenstronomy}} \citep{lenstronomy1, lenstronomy2}.
The process followed several steps. We begin by drawing a random foreground galaxy with lens redshift $z_{lens}$ and velocity dispersion $\sigma_V$ from the deflector sample. From the set of background sources, we select a random galaxy with the source redshift $z_{src} > z_{lens}$. 
We approximate the potential of the deflector with a singular isothermal sphere profile (SIS) which is set by the velocity dispersion $\sigma_V$:
\begin{equation}
    \displaystyle \rho(r) = \frac{\sigma_V^2}{2\pi G r^2}
\end{equation}
In addition to the the deflector potential, we also add the external shear due to the large-scale structure.
We compute the Einstein radius, $\theta_E$, of the system and only keep lenses with $0.75" < \theta_E < 3.0"$. The lower limit is chosen to exceed the seeing in HSC \textit{i} band, and the upper limit is set based on the size of the cutout (64 pix  = $10.2"$) to ensure that the resulting lens fits well within our 64$\times$64 image. We place the background source at a random location in the source plane with an offset ($\Delta X$, $\Delta Y$) drawn from the uniform distribution $[-0.3 \theta_E$, $0.3 \theta_E]$. For the case of SIS, it can be shown that multiple images are produced when the source position is smaller than Einstein radius of the lens. 

\begin{figure}
    \centering
    \includegraphics[width=\columnwidth]{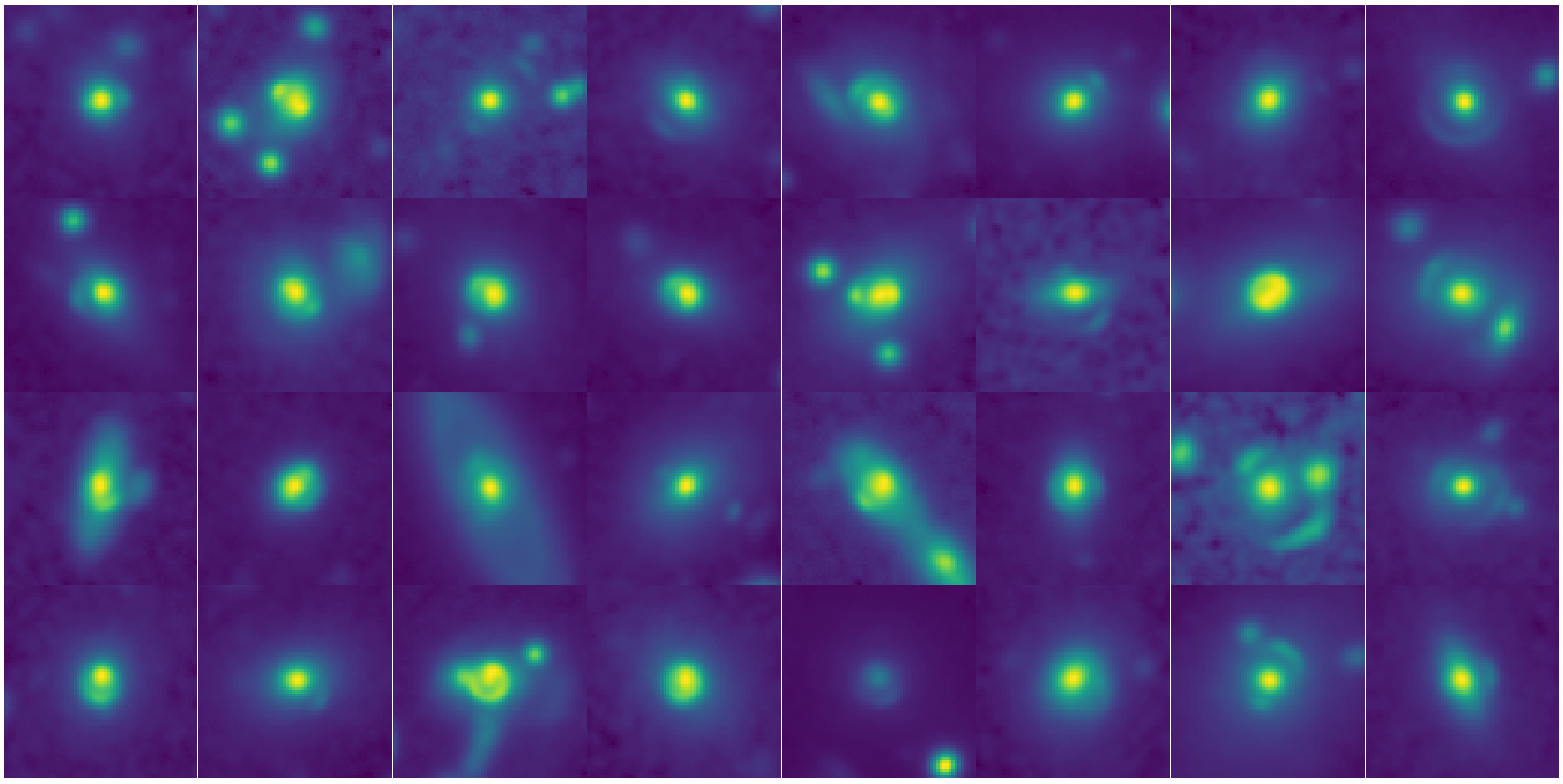}
    \caption{Examples of mock lenses simulated with lenstronomy.}
    \label{fig:sim_examples}
\end{figure}

The image of the background source might contain multiple sources. To remove contaminants and compute the lensed signal only for the central source, we convolve the source image with a Gaussian filter fitted to the central source. 
We compute the distorted light profile of the background source in each filter with the \texttt{lenstronomy} module \texttt{ImSim}. We inspect the resulting image and compare the brightest pixel of the lensed light profile with the corresponding pixel in the deflector image to discard overly faint lenses from the dataset. If the lensed signal is not bright enough, we boost the magnitude of the background source in each filter by 0.5 and try to simulate the lens again. We select a different source if we could not obtain a bright enough lensed image after the total magnitude boost of 5 mag. We coadd the successful lensed image with the deflector image and apply the background noise simulated with \texttt{lenstronomy} for the PSF and observing conditions (e.g. seeing and sky brightness) typical for HSC. Fig.~\ref{fig:sim_examples} shows randomly selected examples of mock lenses in the \textit{i} band.

\begin{figure*}
    \centering
    \includegraphics[width=0.48\textwidth]{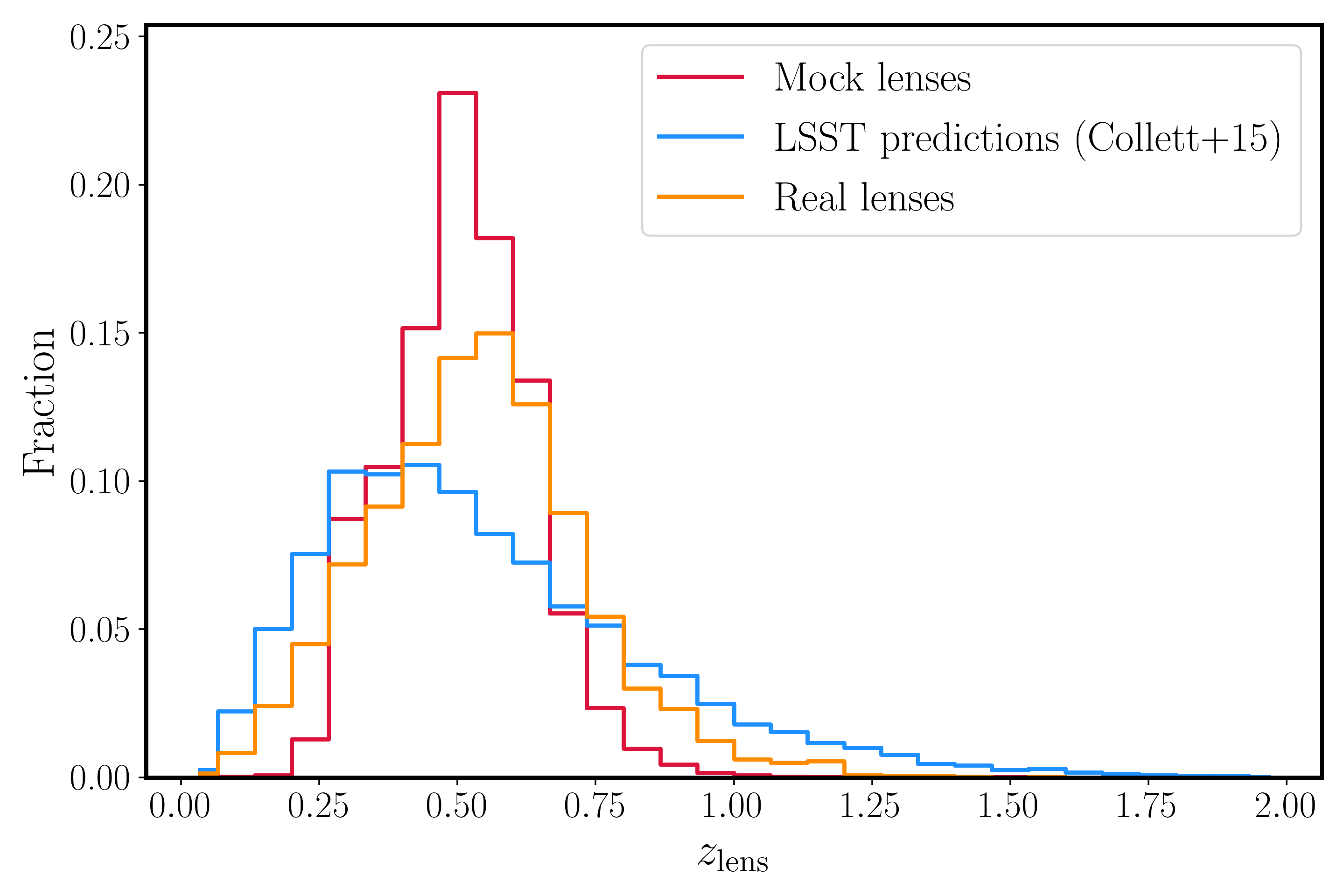}
    \includegraphics[width=0.48\textwidth]{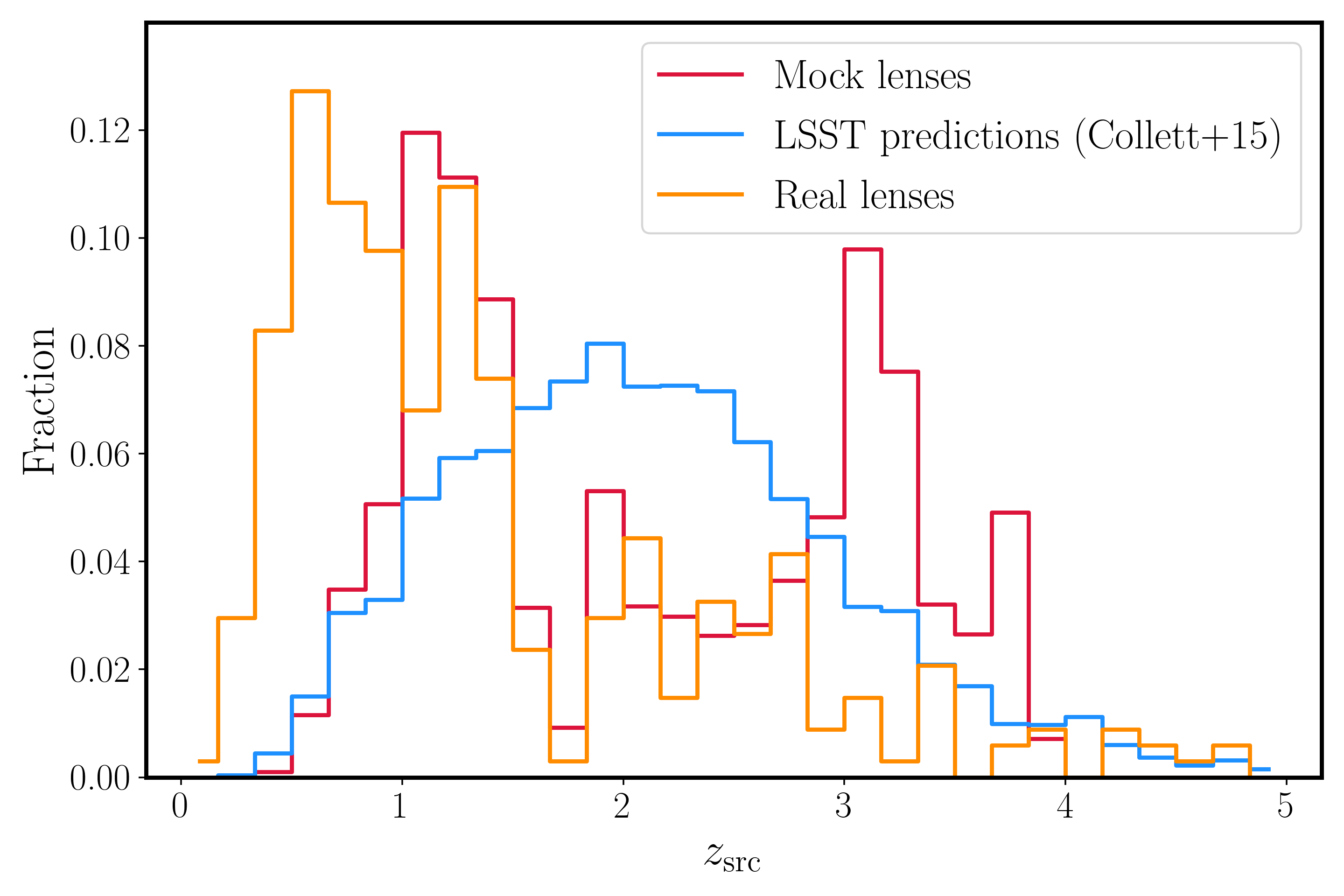}
    \includegraphics[width=0.48\textwidth]{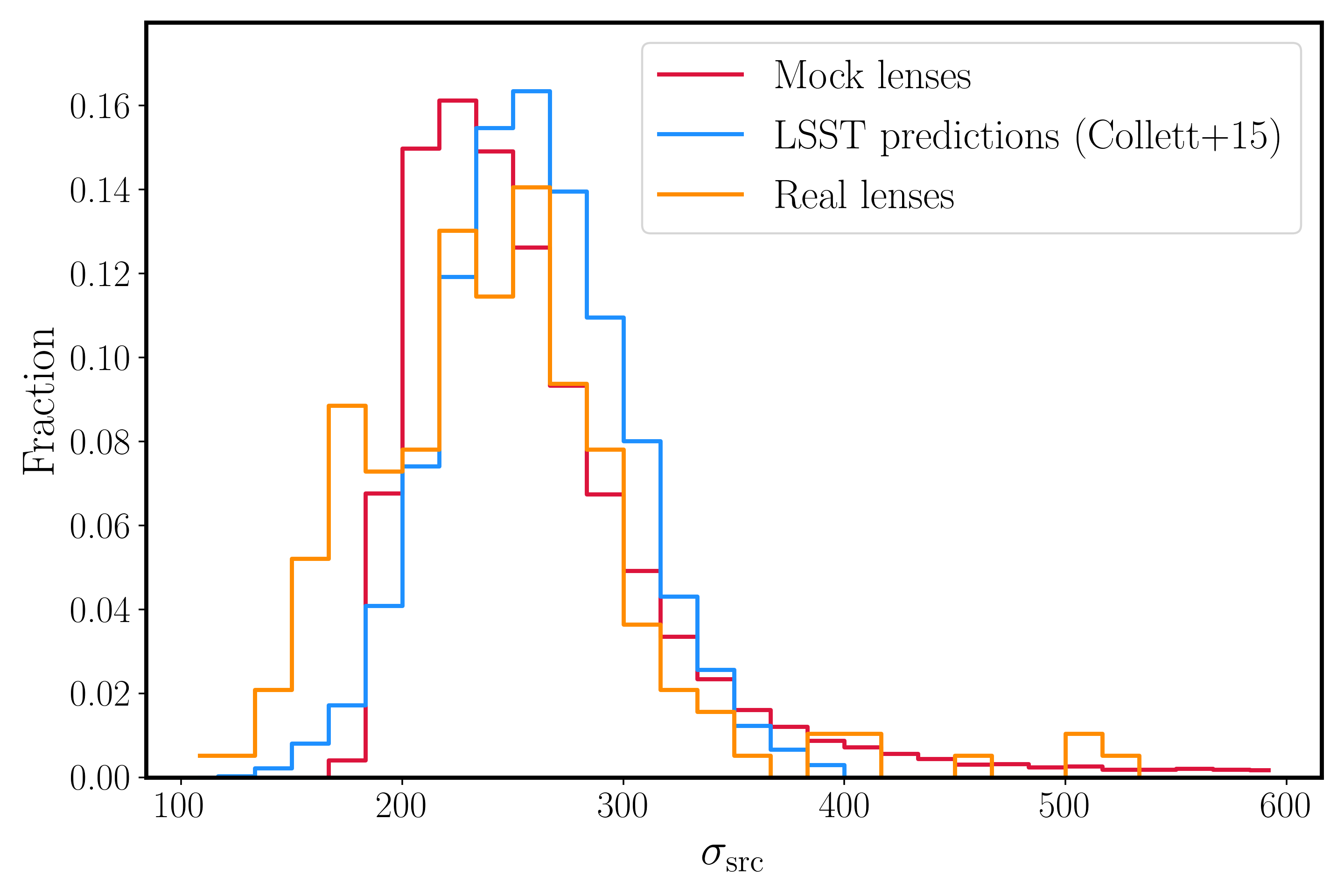}
    \includegraphics[width=0.48\textwidth]{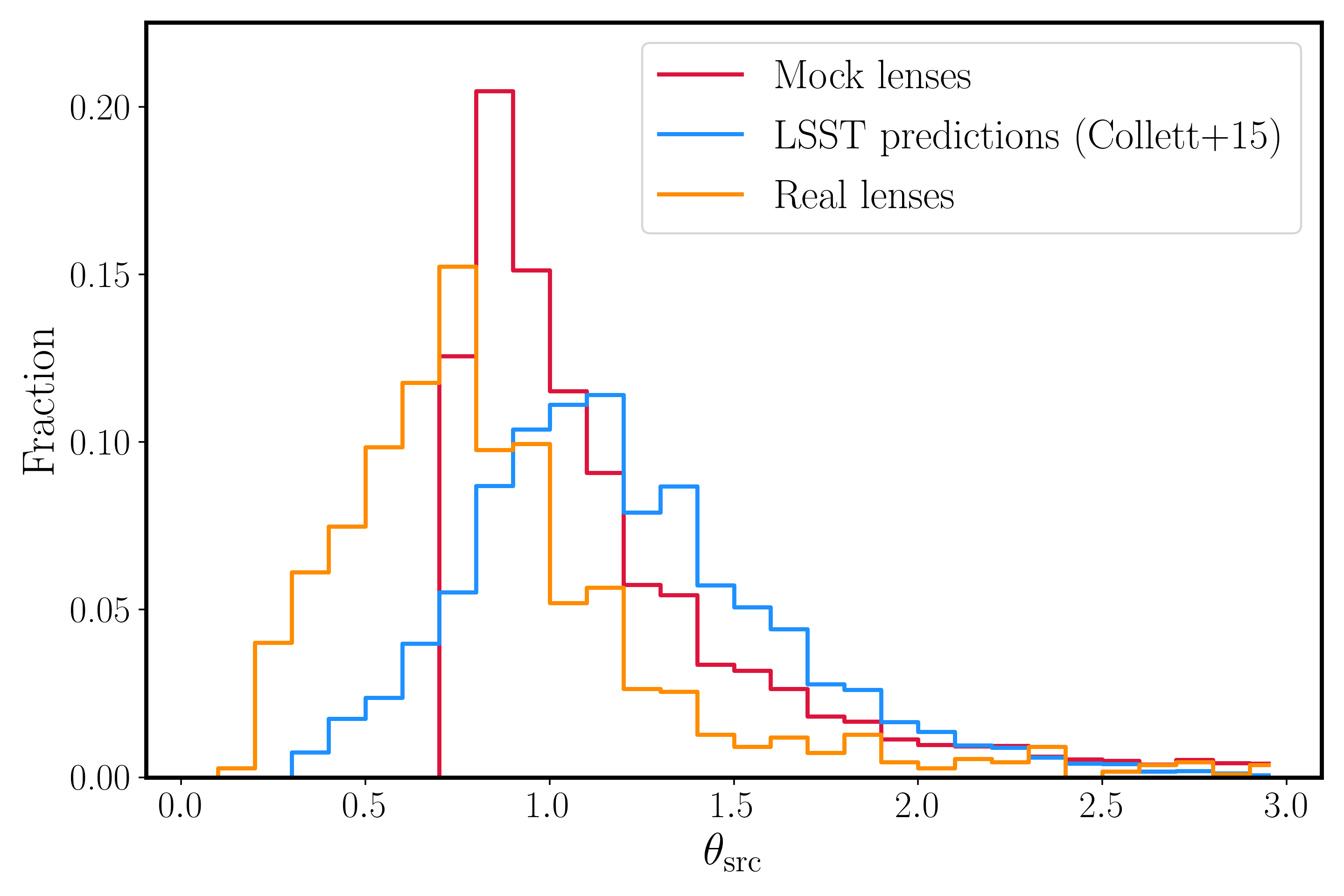}
    \caption{Distribution for the simulated lens systems, real lens systems, and LSST predictions, for lens redshift (top-left), source redshift (top-right), velocity dispersion (bottom-left), and Einstein radii (bottom-right).}
    \label{fig:zlens}
\end{figure*}

Panels in Fig.~\ref{fig:zlens} compare the distributions of different properties of the simulated lens population, predictions for LSST from \citet{Collett2015}, and a subsample of real lens candidates, for which these properties are available in the literature. 
The top two rows compare the distributions of redshifts for the deflectors and the background sources. While the distribution of lens redshifts is similar in shape among the three datasets and displays a peak at $z_{lens} \approx 0.5$--$0.75$, the distribution of redshifts for background sources differs significantly for the mock lenses and display a bimodal behavior, that results from the two peaks in the redshift distribution of the background sources from \citet{Inami}. The lower left panel compares the distributions of velocity dispersion for deflector galaxies, considered as a proxy for galactic mass. All three samples are dominated by lenses with $\sigma_V \approx 250$-$300~\mathrm{km/s}$. The lower right panel compares the Einstein radii of the systems and shows a peak in all three distributions at $\theta_E \approx 0.75"$--$1.25"$, with the LSST-predicted population displaying slightly larger radii. The sharp cutoff at $\theta_{E, \mathrm{mock}} = 0.75"$ comes from the threshold imposed in the simulation.

\subsection{Real lenses}
To construct positive examples in the target dataset, we compiled a comprehensive list of known lens systems and lens candidates. These systems were discovered in previous campaigns through various methods, including both machine learning and traditional approaches. We added all sources from the Master Lens Database\footnote{\url{https://test.masterlens.org/index.php}} (version July 2021), SuGOHI Candidate List\footnote{\url{https://www-utap.phys.s.u-tokyo.ac.jp/~oguri/sugohi/}}, and other published catalogs: \citet{Canameras_HSC, Canameras_panstarrs, Diehl, Garvin, Huang2020, Huang2021, Jacobs2019, petrillo, Rojas, Shu, Stein22, Li, Storfer}. We performed an internal cross-match to remove duplicate sources and obtained a final list of around 12000 galaxy-galaxy systems that were identified across multiple surveys. We cross-matched the full catalog with HSC PDR2 Wide layer and extracted cutouts in the \textit{g}, \textit{r}, and \textit{i} bands. 
We performed a visual inspection of extracted images and excluded group-scale lenses or objects with barely visible or unclear lensed features, which is possible when the lens was discovered in a survey with higher resolution or depth than HSC.
The final sample of real lenses included 1954 objects. We set aside 200 randomly selected lenses for the test set and used the rest in the target dataset.  

In summary, we prepared the following datasets:
\begin{enumerate}
    \item A balanced source dataset consisting of 30 000 mock lenses and 30 000 non-lensed galaxies.
    \item A balanced target dataset comprising 1754 real lenses and 1754 non-lenses.
    \item An unbalanced real train dataset comprising 1754 real lenses (from target) and 31 754 non-lenses (from combining source and target non-lensed sources)
    \item A test dataset containing 200 real lenses and 20 000 non-lenses to reflect the unbalanced nature of real-world datasets. 
\end{enumerate}

We explore domain adaptation methods in two settings. For unsupervised domain adaptation, we compare it against a model trained solely on the source dataset with simulated lenses and applied to the test dataset with real lenses in naive way, i.e. without any domain adaptation. The supervised domain adaptation is compared against a model trained on the unbalanced real dataset. In both cases, we run inference on the test dataset containing only real data.

\section{Network architecture and training} \label{arch}

We compare the performance of three domain adaptation methods described in Sec.~\ref{DA}. While these methods differ in their training approaches, they share common components: an encoder that maps input images to lower-dimensional representations, and a classifier that outputs binary classes for these representations.

For the encoder, we explored two different architectures: the residual neural network \citep[ResNet,][]{He_resnet} and the equivariant neural network \citep[ENN,][]{Weiler_enn}.
ResNet is a type of convolutional neural networks that was developed to overcome the problem of degrading accuracy with an increasing number of layers in earlier models such as AlexNet or VGG. Its key component is the residual block with skip connections. In a residual block, the input is propagated through the layers without being changed, so the block is learning the difference between the original function and an identity transformation. Since their introduction, ResNets have gained popularity in image recognition tasks; specifically in the field of gravitational lens detection, a ResNet-based model won the gravitational lens finding challenge \citep{metcalf_challenge, Lanusse2018}.

We based the encoder on the ResNet-18 architecture. The encoder's core consists of four residual blocks, each containing two convolutional layers with batch normalization and ReLU activations. We modified the standard ResNet-18 architecture by replacing its final fully connected layer with a custom module: a dropout layer for regularization, followed by a linear layer that outputs a 256-dimensional embedding vector of the input image.

In addition to ResNet, we explored equivariant neural networks (ENN), which are designed to preserve inherent symmetries in the data through their architecture. For gravitational lenses, which often display rotational and reflectional symmetries, the use of ENNs can lead to more efficient learning, better generalization, and a reduced need for data augmentation. For example, in \citet{deeplense_DA} domain adaptation with an ENN-based encoder showed superior performance in the task of classifying signatures of dark matter substructures in simulated strong gravitational lenses. 

We used the \texttt{e2cnn} \citep{e2cnn} package and implemented an ENN with the dihedral group D4, which includes the identity transformation, rotations by $\pm \pi/2$ and $\pi$, and horizontal/vertical reflections. Our model comprises six equivariant convolutional blocks, each composed of a convolutional layer, a batch normalization layer, and a ReLU activation function. The final layer also outputs a 256-dimensional representation.

For the classification task, both encoders are followed by a simple classifier network, consisting of two fully connected layers for the binary classification.

We used data augmentation with rotations and horizontal/vertical flips. While augmentation is not required for ENNs due to symmetries being incorporated in the structure of the network, it turned out to be crucial for the successful training of the ResNet-based model.

We used the Adam optimizer \citep{Adam} to minimize losses and 1-cycle scheduler to adjust the learning rate. We trained both ENN and ResNet-based networks for 100 epochs with a patience of 5 epochs, so the training stops if the validation loss of the model does not improve for 5 consecutive epochs.

After the source encoder and the classifier are trained on the source dataset in a supervised way, we continue with the domain adaptation training.

In the ADDA method, the training step is split into two parts. First, we update a discriminator which operates on the embeddings and define whether they come from a source or target dataset. In our case, the discriminator is implemented as a simple three-layer network. Next we update a target encoder, initialized with the same weights as the source encoder, using inverted labels. For the ResNet-based model, we update encoder 5 times for each discriminator step for more stable training. 

For the WDGRL approach, we train a critic network and update the pre-trained encoder and classifier. The goal of the critic network is to estimate the Wasserstein distance between the representations of source and target images. For each mini-batch, we first train the critic for \texttt{n\_critic} steps and then train the encoder and classifier with cross-entropy loss for the classifier and Wasserstein distance as the loss for the encoder. 

For the supervised domain adaptation, we also update the pre-trained classifier and encoder; however we use a contrastive semantic alignment (CSA) loss. In this case, the most important hyperparameters are the balancing coefficient $\alpha$, which determines the relative contribution of classifier loss and CSA loss, and the margin $m$, which sets the minimal separation distance for the examples from different domains and different class labels. 

In all cases we use the Adam optimizer to optimize the losses. 
All hyperparameters were fine-tuned via grid search and are listed in Table \ref{tab:hyperparameters}.
The models were trained on the high-performance computing cluster at The University of Alabama.

\begin{deluxetable}{lccc}
\tablecaption{Hyperparameters used for training DA algorithms}
\label{tab:hyperparameters}
\tablenum{1}
\tablehead{
\colhead{Method} & \colhead{Parameters} & \colhead{ResNet-based} & \colhead{ENN-based}
}
\startdata
\multirow{4}{*}{Supervised} & learning rate (lr) & 6e-4 & 1e-5 \\
                            & scheduler lr & 1e-4 & 0.002 \\
                            & weight decay & 0.001 & 1e-6 \\
                            & batch size & 128 & 128 \\
\hline
\multirow{3}{*}{ADDA}       & target encoder lr & 1e-6 & 1e-6 \\
                            & discriminator lr & 5e-5 & 1e-5 \\
                            & epochs & 20 & 20 \\
\hline
\multirow{6}{*}{WDGRL}      & domain critic lr & 5e-5 & 1e-4 \\
                            & classifier lr & 1e-5 & 1e-4 \\
                            & encoder lr & 1e-3 & 1e-4 \\
                            & n\_critic & 5 & 5 \\
                            & $\gamma$ & 0.3 & 1 \\
                            & epochs & 20 & 20 \\
\hline
\multirow{5}{*}{SDA}        & encoder lr & 1e-4 & 1e-3 \\
                            & classifier lr & 1e-4 & 1e-3 \\
                            & $\alpha$ & 0.3 & 0.75 \\
                            & $m$ & 8 & 15 \\
                            & epochs & 40 & 20 \\
\enddata
\end{deluxetable}

\begin{deluxetable}{lccccc}
\tablecaption{Results for unsupervised domain adaptation algorithms}
\label{tab:results_lenses_unsup}
\tablenum{2}
\tablehead{
\colhead{Method} & \colhead{AUROC} & \colhead{$\mathrm{TPR}_{1\%}$} & \colhead{$\mathrm{F1}_{1\%}$} & \colhead{$\mathrm{TPR}_{0.1\%}$} & \colhead{$\mathrm{F1}_{0.1\%}$}
}
\startdata
ENN (naive) & 0.921          & 0.354          & 0.303           & 0.             & 0. \\
ADDA+ENN        & \textbf{0.942} & 0.462          & 0.377           & 0.             & 0. \\
WDGRL+ENN       & 0.940          & \textbf{0.528} & \textbf{0.416}  & \textbf{0.241} & \textbf{0.360} \\
\hline
ResNet (naive) & 0.858          & 0.262          & 0.231           & 0.046 & 0.080 \\
ADDA+ResNet           & 0.886          & 0.236          & 0.210           & 0.036          & 0.061 \\
WDGRL+ResNet          & \textbf{0.903} & \textbf{0.277} & \textbf{0.242}  & \textbf{0.072}             & \textbf{0.122} \\
\enddata
\end{deluxetable}

\section{Results} \label{res}
To compare the efficiency of domain adaptation techniques for lens finding, we use the Receiver Operation characteristic Curve (ROC) curve as our main metric. The ROC curve illustrates how the false positive rate (FPR) and true positive rate (TPR) change with the variation of the classification threshold. TPR, also known as sensitivity or recall, is defined as the number of correctly classified lenses relative to the total number of lenses in the test dataset: 
\begin{equation}
    \mathrm{TPR = \frac{TP}{TP+FN}}
\end{equation}
FPR is the ratio of incorrectly classified non-lensed objects to the total number of non-lenses:
\begin{equation}
    \mathrm{FPR = \frac{FP}{TN+FP}}
\end{equation}
For an ideal classifier, FPR is close to 0 when the TPR is close to 1. For a random classifier, FPR equals TPR for all thresholds. The model that performs better has higher area under the ROC curve (AUROC).

Besides the AUROC, it is also particularly interesting to compare the behaviour of the ROC curve at low FPRs; we quantify it by measuring $\mathrm{TPR}_{1\%}$ and $\mathrm{TPR}_{0.1\%}$ -- true positive rates at the thresholds that provide false positive rate of 1\% and 0.1\% respectively. In practice, all objects classified as positive examples by automated methods are visually validated. For the purpose of saving time in human inspection, the ideal algorithm should have a very low contamination fraction while maintaining a high number of identified lenses. 

\begin{figure*}
    \centering
    \includegraphics[width=0.49\textwidth]{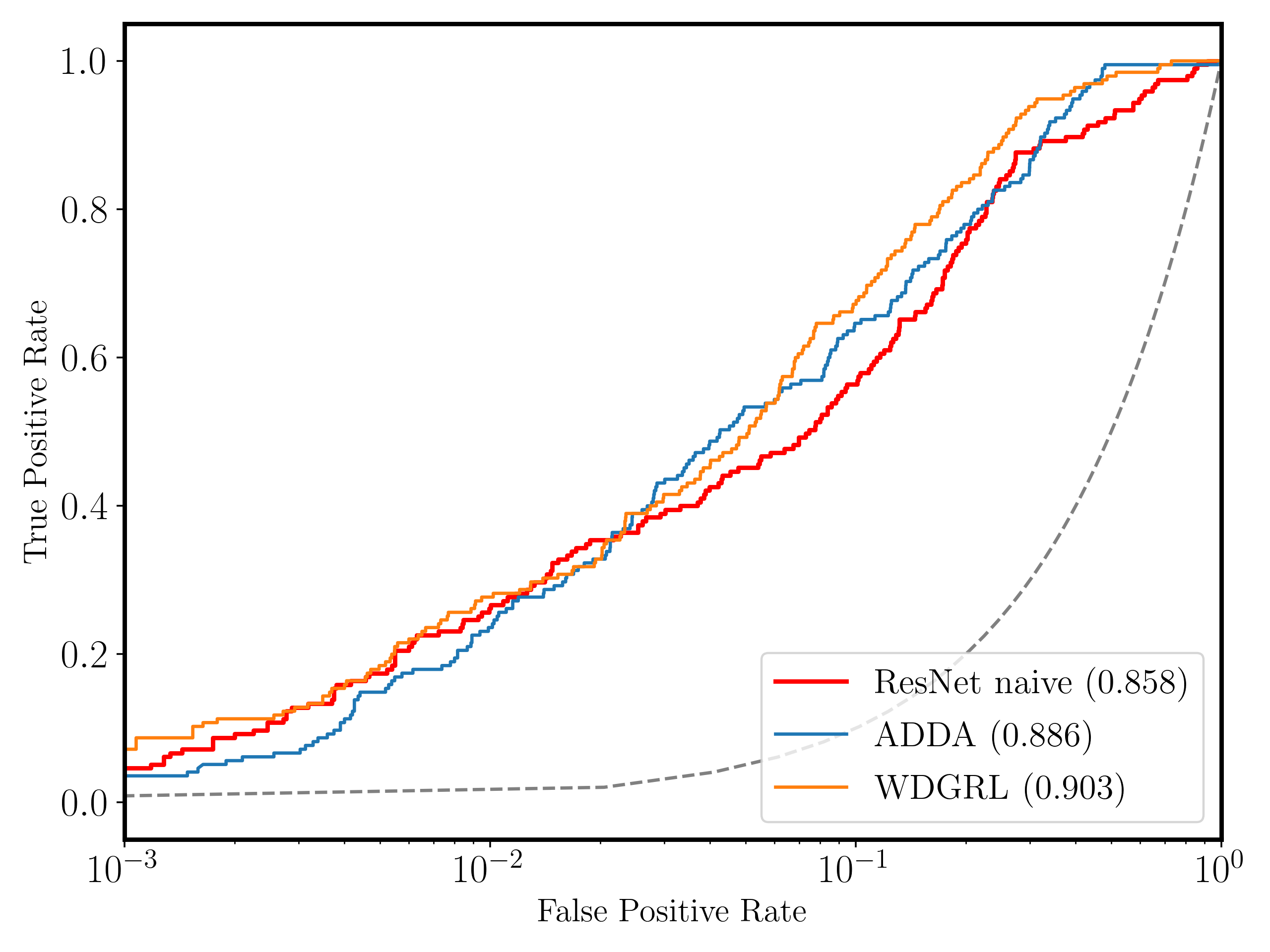}
    \includegraphics[width=0.49\textwidth]{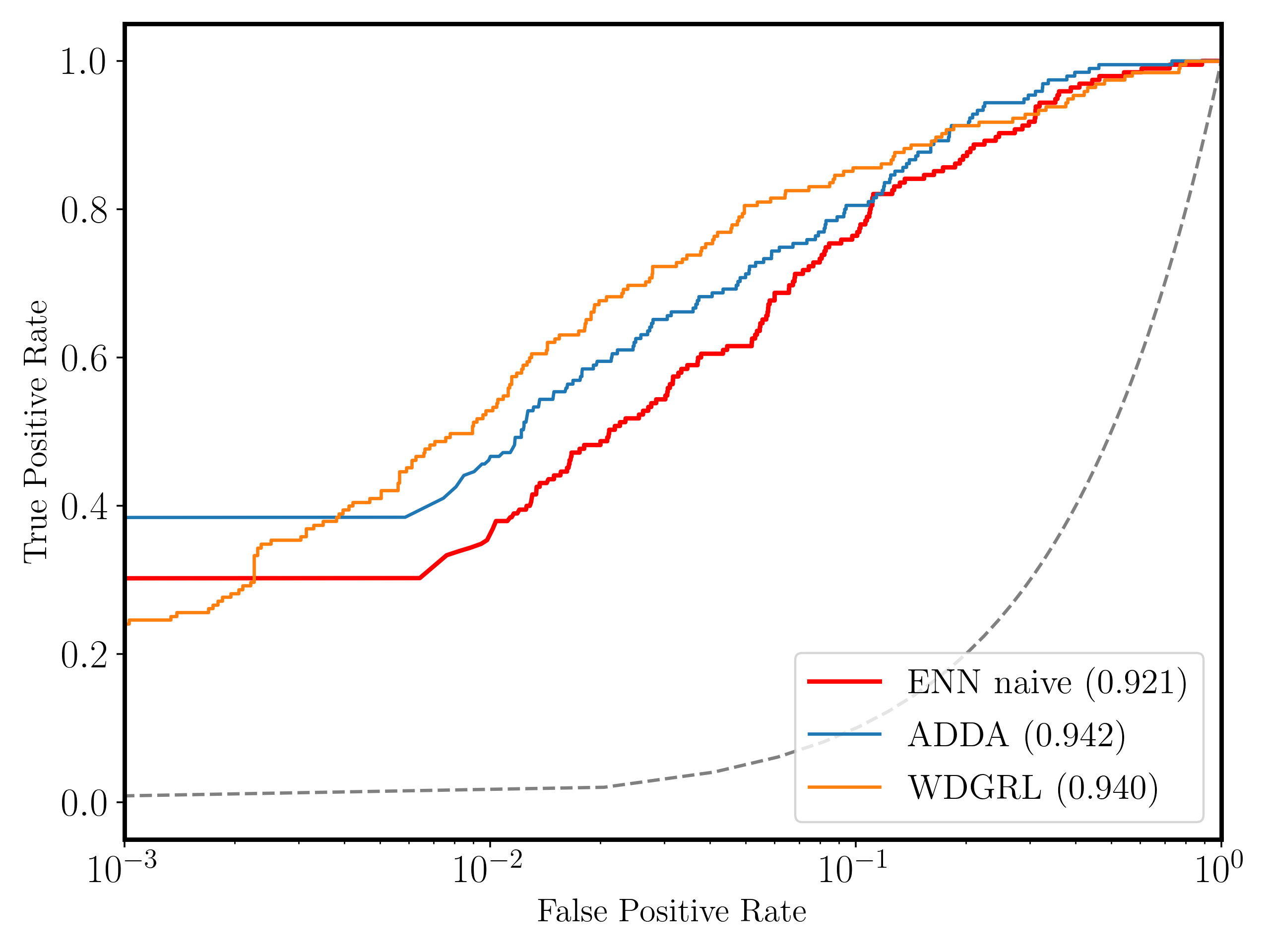}
    \caption{ROC curves for ResNet-18-based (left) and ENN-based (right) algorithms with unsupervised domain adaptation (blue and orange curves) compared to the model trained on a source dataset and applied to the test dataset without domain adaptation (red curve). Dashed line represents ROC curve of a random classifier. }
    \label{fig:roc_unsup}
\end{figure*}

Fig. \ref{fig:roc_unsup} displays ROC curves for unsupervised domain adaptation methods with a ResNet-based encoder on the left panel and an ENN-based encoder on the right, evaluated on the imbalanced test dataset with 200 lenses and 20 000 non-lenses. 
For the ResNet, both ADDA and WDGRL result in an overall higher AUROC; however, the increase in performance is rather small, and at low false positive rates the recall of all three algorithms is similar (see Table~\ref{tab:results_lenses_unsup}).  
Compared to ResNet, the ENN-based model works much better even without domain adaptation (AUROC score 0.921 vs 0.858) and improvement from application of domain adaptation is more prominent, especially at low false positive rates: $\mathrm{TPR}_{1\%} = 0.354$ for the naive approach, 0.462 for ADDA, and 0.528 for WDGRL. In both cases, the WDGRL method outperforms ADDA.

\begin{figure*}
    \centering
    \includegraphics[width=0.49\textwidth]{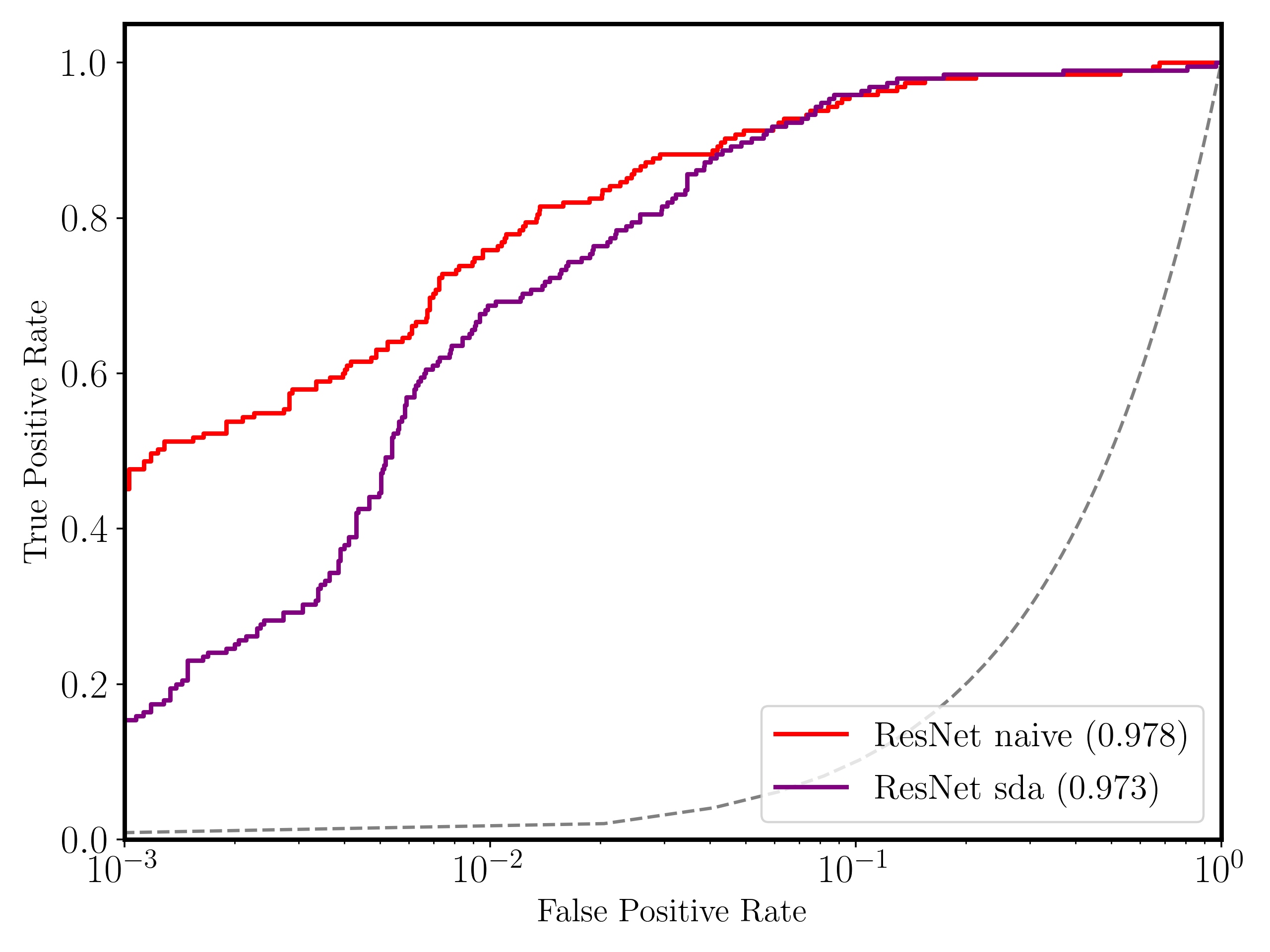}
    \includegraphics[width=0.49\textwidth]{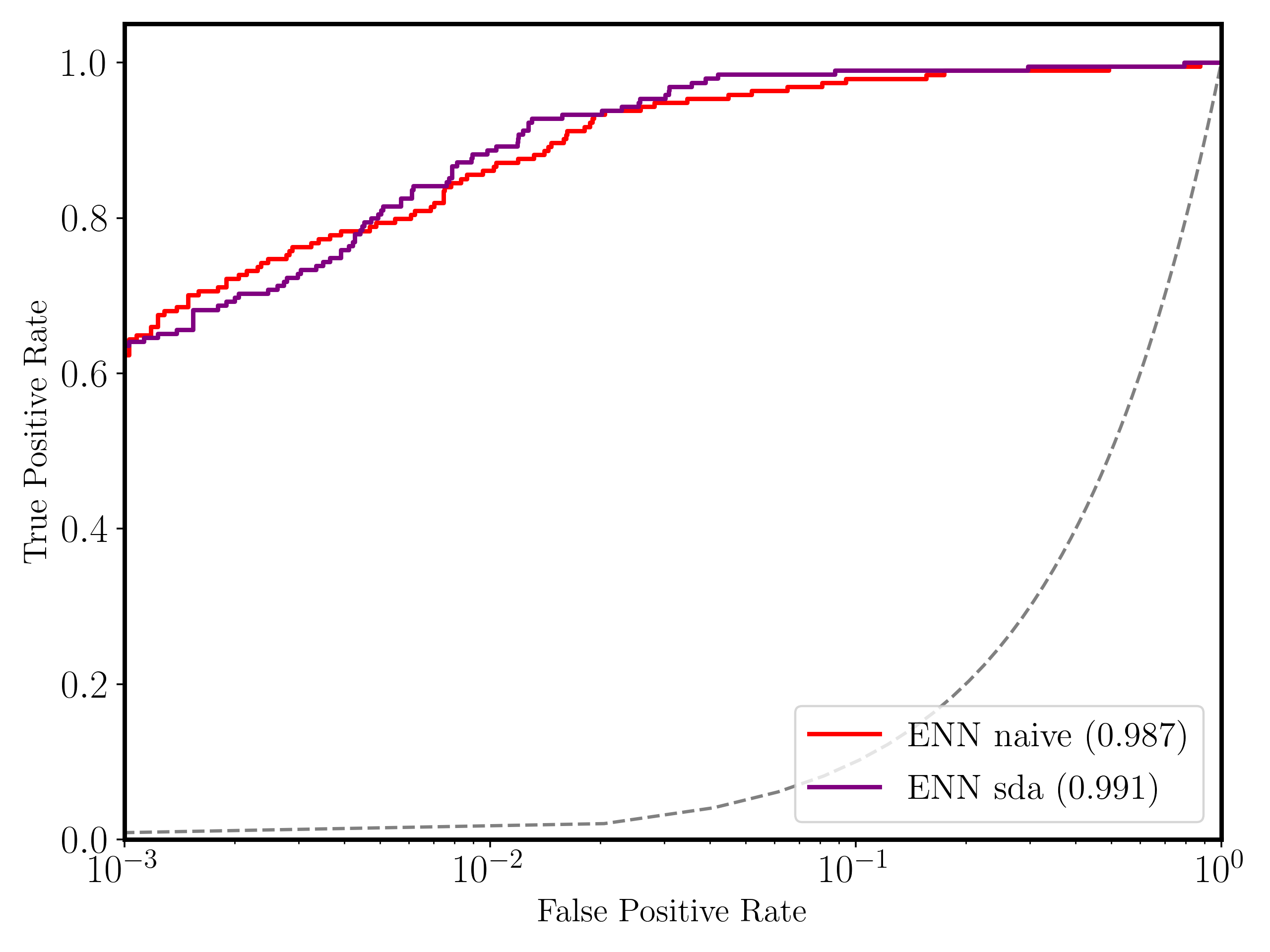}
    \caption{ROC curves for ResNet-18-based (left) and ENN-based (right) algorithms with supervised domain adaptation (blue curve) compared to the model trained on a source dataset and applied to the test dataset without domain adaptation (red curve). Dashed line represents ROC curve of a random classifier.}
    \label{fig:roc_sup}
\end{figure*}

Fig.~\ref{fig:roc_sup} demonstrates the results for the supervised domain adaptation. In this setting, we compared the adapted model (blue curve) with a model trained on an unbalanced dataset containing only observational images of lensed and non-lensed sources (red curve).
The supervised domain adaptation in the case of ENN-based encoder gives some improvement, however it is rather small (AUROC score 0.987 vs 0.991). For the ResNet-based classifier, SDA performed even worse than naive approach.
One interesting result is that the model, trained on real sources, significantly outperform the classifiers trained on simulations (compare red curves on Fig.~\ref{fig:roc_unsup} and \ref{fig:roc_sup}), with AUROC for ENN improving from 0.921 to 0.987 and for ResNet -- from 0.858 to 0.978, even though the number of real lenses is $\sim 15$ times smaller than the number of mock lenses. While the unbalanced proportion of the training dataset is in better agreement with the unbalanced nature of the test dataset and of real world data, it is usually considered a good practice to balance training dataset to make sure that the model is able to see a sufficient number of examples from all classes and learn relevant features. The fact, that the model trained on a smaller set of real data performs better than the model trained on a larger dataset with simulations highlights that simulations are not perfect representation of real data. Another explanation might come from the fact that majority of real lens candidates that consitute training and test set were found with machine learning algorithms and might represent subsample of objects with similar properties, for example, in terms of relative brightness and size of lens features. Therefore, in future work it would be interesting to apply both models to the larger sample of unseen data and compare properties of the identified lens candidates.

\begin{deluxetable}{lccccc}
\tablecaption{Results for supervised domain adaptation algorithm}
\label{tab:results_lenses_sup}
\tablenum{3}
\tablehead{
\colhead{Method} & \colhead{AUROC} & \colhead{$\mathrm{TPR}_{1\%}$} & \colhead{$\mathrm{F1}_{1\%}$} & \colhead{$\mathrm{TPR}_{0.1\%}$} & \colhead{$\mathrm{F1}_{0.1\%}$}
}
\startdata
ENN (naive) & 0.987 & 0.861 & 0.597 & 0.624 & 0.725 \\
SDA+ENN         & \textbf{0.991} & \textbf{0.887} & \textbf{0.636} & \textbf{0.734} & \textbf{0.537} \\
\hline
ResNet (naive) & \textbf{0.978} & \textbf{0.759} & \textbf{0.542} & \textbf{0.451} & \textbf{0.583} \\
SDA+ENN            & 0.973 & 0.687 & 0.506  & 0.154 & 0.245 \\
\enddata
\end{deluxetable}

\subsection{Spiral contaminants}
One of the major challenges in reducing the number of false positives is the correct classification between strong lenses and objects with similar arc-like features, such as spiral galaxies, ring galaxies, or galaxy mergers. While these objects are not as rare as gravitational lenses, a randomly sampled subset of non-lenses typically includes only a small fraction of such objects. Consequently, a trained classifier can easily become confused by these similar-looking patterns.

To explore the potential of domain adaptation in mitigating this problem, we conducted the following experiment. Using the catalog of spiral galaxies from \citet{Tadaki20}, we randomly drew 1754 spiral galaxies (to match the number of real lenses) from a subsample of spirals with redshift $z > 0.4$ to align with the distribution of redshifts of non-lenses and deflectors. We used our best performing model, which uses an ENN-based encoder combined with supervised domain adaptation, and retrained it while including the spiral subsample at different stages of the training. We obtained three models: 1) with spiral galaxies added to the source dataset; 2) with spiral galaxies added to the target dataset; and 3) with spiral galaxies split equally between the source and the target dataset. We compared the SDA model with the model trained solely on the unbalanced real dataset, to which we also added the same spiral galaxies.
We also constructed a modified test dataset: to the set of 200 real lenses and 20 000 randomly sampled non-lenses we added 2000 spirals from the same catalog.

\begin{figure}
    \centering
    \includegraphics[width=\columnwidth]{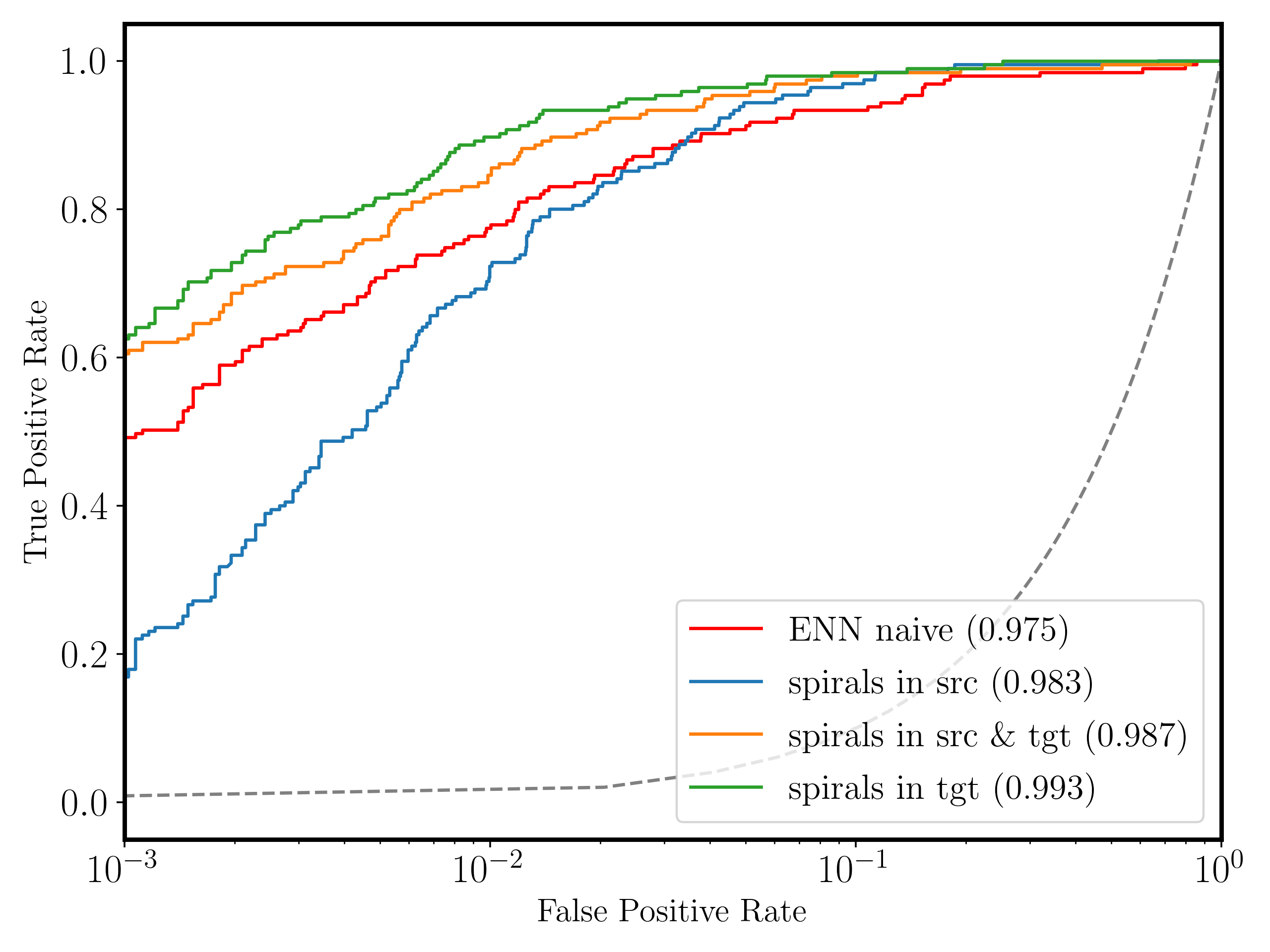}
    \caption{ROC curves for ENN-based algorithm tested on a dataset with spiral contaminants. Red curve represents the supervised classifier without SDA trained on the dataset containing spiral galaxies among non-lenses. The rest of the curves represent the results for SDA model with spiral galaxies added only to the source dataset (blue), only to the target dataset (green), or splitted between the source and target (orange). Dashed line represents ROC curve for a random classifier.}
    \label{fig:spirals}
\end{figure}

The results of the evaluation on the modified test dataset are shown in Fig.~\ref{fig:spirals}. The naive supervised classifier has the lowest AUROC, however it shows much higher sensitivity at low FPR than the SDA model that has seen spiral galaxies only in the source dataset.  The SDA model, that was trained on a target dataset containing all spiral galaxies, shows the best result, both in terms of AUROC and $\mathrm{TPR}_{1\%}$ and $\mathrm{TPR}_{0.1\%}$. While SDA model trained in the setting where sample of spirals was split between the source and target dataset shows the second best result. The summary of results is listed in Table~\ref{tab:spirals}.

\begin{deluxetable}{lccccc}
\tablecaption{Results for supervised domain adaptation algorithm trained on a dataset with spiral galaxies.}
\label{tab:spirals}
\tablenum{4}
\tablehead{
\colhead{Method} & \colhead{AUROC} & \colhead{$\mathrm{TPR}_{1\%}$} & \colhead{$\mathrm{F1}_{1\%}$} & \colhead{$\mathrm{TPR}_{0.1\%}$} & \colhead{$\mathrm{F1}_{0.1\%}$}
}
\startdata
ENN (naive) & 0.975 & 0.774 & 0.539 & 0.492 & 0.613 \\
spirals in src & 0.983 & 0.723 & 0.511 & 0.169 & 0.265 \\
spirals in tgt & \textbf{0.993} & \textbf{0.897} & \textbf{0.586} & \textbf{0.626} & \textbf{0.722} \\
spirals in src \& tgt  & 0.987 & 0.846 & 0.574  & 0.605 & 0.707 \\
\enddata
\end{deluxetable}

This experiment demonstrates that supervised domain adaptation with a carefully constructed target dataset is able to improve the classifier's ability to differentiate between lenses and similar-looking objects, such as spiral galaxies. This approach could be extended to other common contaminants such as ring galaxies or mergers, potentially leading to more robust and reliable automated lens detection. Future work could also explore the integration of superresolution techniques \citep[e.g.][]{difflense} to enhance the quality of input images, potentially improving the performance of both domain adaptation methods and traditional classifiers.

\section{Discussion \& Conclusion} \label{disc}
In this work, we investigate whether domain adaptation is able to enhance the performance and reliability of CNN-based gravitational lens detection algorithms. Due to the limited sample of known lenses, deep learning models are typically trained on simulated strong lenses, which can lead to degraded performance on real data. The shift between training and test domains can be alleviated with domain adaptation techniques. 

We used simulated mock lenses as positive examples in the source dataset and observations of lens candidates from HSC SSP PDR2 Wide layer as constituents in the target dataset. We explored two neural network architectures, ResNet-18 and Equivariant Neural Network, for the encoder and found that ENNs are more successful in identifying lenses. 

For unsupervised domain adaptation, we implemented two methods, ADDA and WDGRL. The WDGRL approach resulted in the largest increase of ROC score compared to the naive inference of a classifier trained with mock lenses on real data. 
For the supervised domain adaptation, we compared the performance of the adapted algorithm with a model trained on an unbalanced dataset containing purely observational data. Somewhat surprisingly, despite the relatively small number of lenses in the training sample (1754 real lenses vs 31754 mock lenses in the source dataset), the model showed superior performance. Supervised domain adaptation provided only marginal improvement in the case of the ENN-based algorithm, and did not yield any improvement for the ResNet-based model. 
  
In the common setting for lens searches, algorithms are applied to large datasets ($10^6-10^7$ objects), and objects with a classification score above a predefined threshold are visually inspected by a team of experts. A higher threshold leads to increased sample purity at the cost of reduced completeness, and vice versa. This trade-off means that human experts must examine a large number of lens candidates. For example, in \citet{Jaelani2023}, 20 241 sources with a score higher than 0.9 were visually inspected, first reduced to 1522 cutouts, and then to 43 definite and 269 probable lenses. Upcoming surveys are expected to contain $\mathcal{O}(10^5)$ lenses, which makes the task of reducing false positives particularly important. In our experiment we saw that the improvement in recall from the combination of ENN-based algorithm and WDGRL domain adaptation method is particularly noticeable for low false positive rates. 

One of the challenges of the lens finding task is the correct classification between lenses and similar-looking contaminants, such as ring galaxies or spiral galaxies. In the supervised domain adaptation setting, we show that an algorithm trained on a target dataset that includes a relatively small number of spirals is able to enhance the model's ability to distinguish between lenses and spiral galaxies. The model's performance increases with the number of spiral galaxies in the target dataset. However, adding spirals only to the source dataset leads to performance degradation. 

Domain adaptation techniques can be seamlessly integrated into lens finding pipelines and easily combined with other methods. Our results emphasize that DA methods should be considered an essential component in future lens finding campaigns.

\section{Acknowledgments}
We acknowledge useful conversations with Stephon Alexander. H. P. and S. G. were supported in part by U.S. National Science Foundation award No. 2108645. P. R. was a participant in the Google Summer of Code 2023 program.  
This material is based upon work supported by the U.S. Department of Energy, Office of Science, Office of High Energy Physics of U.S. Department of Energy under grant Contract Number DE-SC0012567. M. W. T. acknowledges financial support from the Simons Foundation (Grant Number 929255).

\software{Astropy \citep{astropy:2013, astropy:2018, astropy:2022}, SciPy \citep{2020SciPy-NMeth}, Matplotlib \citep{Hunter:2007}, NumPy \citep{harris2020array}, Lenstronomy \citep{lenstronomy1, lenstronomy2}\footnote{https://github.com/lenstronomy/lenstronomy}, pyHalo \citep{pyhalo}\footnote{https://github.com/dangilman/pyHalo}, PyTorch \citep{Ansel_PyTorch_2_Faster_2024}.}

\bibliography{sample631}{}
\bibliographystyle{aasjournal}



\end{document}